\documentstyle[epsf,PASJadd]{PASJ95}
\draft

\newcommand{\vol}{}
\newcommand{\no}{}
\newcommand{\lett}{}
\newcommand{\spage}{}
\newcommand{\rdate}{}
\newcommand{\adate}{}

\begin{document}

\title{Overabundance of Calcium in the young SNR RX J0852$-$4622:
evidence of over-production of $^{44}$Ti}

\author{Hiroshi {\sc Tsunemi} \thanks{\it Department of Earth and Space
Science, Graduate School of Science, Osaka University, 1-1
Machikaneyama, Toyonaka, Osaka 560-0043 JAPAN;
tsunemi@ess.sci.osaka-u.ac.jp, miyata@ess.sci.osaka-u.ac.jp,
jhiraga@ess.sci.osaka-u.ac.jp, akutsu@ess.sci.osaka-u.ac.jp}
\thanks{CREST, Japan Science and Technology Corporation (JST), 2-1-6
Sengen, Tsukuba, Ibaraki 305-0047 JAPAN} , Emi {\sc Miyata}$^{*\dag}$,
Bernd {\sc Aschenbach}\thanks{\it Max-Planck-Institut f\"ur
extraterrestrische Physik, D-85740, Garching, Germany;
bra@xray.mpe.mpg.de}, Junko {\sc Hiraga}$^*$, and Daisuke {\sc Akutsu}$^*$}

\abst{ Recently, COMPTEL has detected $\gamma$-rays of 1157\,keV from
$^{44}$Ti in the direction of the SNR RX J0852$-$4622 (Iyudin et
al. 1998).  Since $^{44}$Ti is a product of explosive nucleosynthesis
and its half lifetime $\tau\sb{1/2}$ is about 60\,yrs, RX J0852$-$4622
must be a young supernova remnant and radiation is dominated by the
ejecta rather than by interstellar matter.  We have detected an X-ray
emission line at $4.1\pm 0.2$ keV which is thought to come from highly
ionized Ca.  The emission line is so far only seen in the north-west
shell region of RX J0852$-$4622.  The X-ray spectrum can be well fitted
with that of thin hot plasma of cosmic abundances except that of Ca,
which is overabundant by a factor of $8 \pm 5$.  Assuming that
most of Ca is $^{44}$Ca, which originates from $^{44}$Ti by
radioactive decay, we estimate a total Ca mass of about $1.1\times
10^{-3}M_\odot$.  Combining the amount of $^{44}$Ca and the observed
flux of the $^{44}$Ti $\gamma$-ray line, the age of RX J0852$-$4622 is
around 1000\,yrs.}

\kword{ISM: individual (RX J0852$-$4622) --- ISM: supernova remnants --- X-Rays: sources --- X-Rays: spectra }

\maketitle
\thispagestyle{headings}

\section{Introduction}

A supernova (SN) explosion is a source of heavy elements in the
Galaxy. The nucleosynthesized material inside the star is dispersed into
interstellar space at relatively high temperature.  In young supernova
remnants (SNRs), there are many X-ray emission lines from highly ionized
elements, which tend to be diluted when mixing with ambient interstellar
matter.

There are two types of young SNRs from a morphological point of view:
shell-like structures and those of center-filled morphology.  There are
also two types of young SNRs from the spectroscopic point of view: thin
thermal emission and power-law type spectra.  The Crab nebula, a young
SNR, shows a center-filled structure with a power-law type spectrum.
There is a neutron star, the Crab pulsar, in its center, which is the
energy source of the SNR.  There are more SNRs showing shell-like
structures with spectra due to optically thin thermal emission,
e.g. Cassiopeia-A, Tycho and Kepler's SNRs.  The shell structure is the
result of the interaction of the shock waves and corresponding
shock-heating of the ambient plasma.  This produces thin thermal
emission including many emission lines.  We note that the SNe of Tycho
and Kepler have been witnessed and recorded in historical documents
whereas there is no record of Cassiopeia-A which is estimated to be born
late in the 17th century.

The remnant of the SN which occurred in AD1006, SN1006, is another young SNR.
It shows a pronounced shell-like structure.  Koyama et
al. (1996) revealed that the spectra from the north-east and
south-west regions of the shell are of power-law type while that from
the interior is thin thermal emission.  They claimed that the shell
region in SN1006 is a site of cosmic ray shock acceleration with
energies of up to 200\,TeV.  Accordingly, the shell-like structure does
not always stand for a thermally radiating SNR.

Cassiopeia-A, one of the young SNRs, is the first source from which
$\gamma$-ray emission lines from $^{44}$Ti have been reported (Iyudin et
al. 1994).  $^{44}$Ti is expected to be produced in the explosive
nucleosynthesis inside the SN which also produces $^{56}$Ni.  $^{44}$Ti
decays into $^{44}$Sc emitting two hard X-ray lines of 68\,keV and
78\,keV with a half lifetime, $\tau\sb{1/2}$, of 60 yrs (Ahmad et
al. 1998).  $^{44}$Sc decays further to $^{44}$Ca emitting a
$\gamma$-ray line at 1157\,keV with $\tau\sb{1/2}$ of 3.9 hours which
has been detected with COMPTEL.  In this way, $^{44}$Ti is converted
into $^{44}$Ca.  Taking into account the short lifetime of $^{44}$Ti,
its detection is direct evidence of the source being a young SNR.  The
discovery of this second celestial source of $^{44}$Ti has been reported
by Iyudin et al. (1998): it is located in the direction of the
south-east part of the Vela SNR where Aschenbach (1998) discovered a new
independent SNR, RX J0852$-$4622, which has a radius of $\simeq
1^{\circ}$.  The distance to this source is not clear while it is on the
way to the Vela SNR that is 250 pc away (Cha et al. 1999).  Based on the
X-ray and $^{44}$Ti data, RX J0852$-$4622 is expected to be born several
hundred yrs ago in a sky region which in principle could have been
observed from China, and therefore some record in a historical document
is expected if the SN was of standard brightness.  However, if it were
sub-luminous the event might have been missed by the contemporaries
(Aschenbach et al. 1999).  So far, no record has been found, which
reminds us of some similarity with Cassiopeia-A.  We note that quite
recently a footprint of a historical SN has been found in terms of
nitrate abundance in Antarctic ice cores (Burgess \& Zuber 1999).
The associated SN would have occurred around 1320 $\pm$ 30 yrs.

We have observed RX J0852$-$4622 with ASCA on December 21--24, 1998 as a
TOO target.  This observation was performed with the two CCD cameras
(SIS) (Yamashita et al. 1997) and the two imaging gas scintillation
proportional counters (GIS) (Makishima et al. 1996) which are placed at
the focal plane of four thin-foil X-ray mirrors (Serlemitsos et al. 1995) on
board the ASCA satellite (Tanaka et al. 1994).  In this paper, we
concentrate on the emission from the north-west bright shell
region.  The entire remnant will be reported elsewhere.

\section{Observation and Data Screening}

The observation was performed in 7 pointings so that we could cover the
entire remnant with the GIS. Figure 1 shows the GIS field of view (FOV)
superimposed on the ROSAT all-sky survey image (Aschenbach et
al. 1999). We should note that we did not cover the south bright shell
region.  The observational log is shown in table 1. We excluded all the
GIS data taken at elevation angle below 5$^\circ$ from the night earth
rim and 25$^\circ$ from the day earth rim, a geomagnetic cutoff rigidity
lower than 6 GeV c$^{-1}$, and the region of the South Atlantic Anomaly
(SAA). We also applied a ``flare-cut'' to maximize the signal-to-noise
ratio described in Ishisaki et al. within the frame work of {\sl
ASCA\_ANL} (Ishisaki Y. et al, 1997, ASCA News letter, No.5 ).  In this
way, the effective exposure time is $\simeq$ 76ksec after the data
screening.  We constructed three kinds of images for the GIS data set: a
total count image, a non-X-ray background image, and a cosmic X-ray
background image. As shown in table 1, the GIS bit assignments are
slightly different from pointing to pointing, the binsize of coordinates
is different. We redistributed events in each binning area with a
Poisson random number to make images having the same dimensions.  The
non-X-ray background image was generated with the H02-sorting method in
{\sl DISPLAY45} (the detailed description of this method is in Ishisaki
(1996)). The image of the cosmic X-ray background was extracted from the
Large Sky Survey (LSS) data observed during the ASCA PV phase (Ueda et
al. 1999). We also extracted the non-X-ray background image from the LSS
data to produce the mean cosmic X-ray background image solely.  Thus, we
extracted background-subtracted GIS images in the energy band of
$0.7-10$ keV as shown in figure 2.  We can find a well-defined shell
structure and a compact point-like source at the geometrical center of
the X-ray shell.

In the case of the SIS data analysis, we excluded all the data taken at
elevation angle below 5$^\circ$ from the night earth rim and 30$^\circ$
from the day earth rim, a geomagnetic cutoff rigidity lower than 6 GeV
c$^{-1}$, the region from 128s before entering the SAA to the end of the
SAA, and the period of 120s at the day-night change of the satellite.
After the screening, the total exposure time is $\simeq$59 ks.  We,
next, manually removed the time region where we could see a sudden
change in the light curves of the corner pixels of X-ray events.  Then,
we removed the hot and flickering pixels and corrected CTI, DFE, Echo
effects, and RDD (T. Dotani et al. 1995, ASCA News Letter 3, 25;
T. Dotani et al. 1997, ASCA Letter News 5, 14) in our data sets. We
employed the calibration file of {\tt sisph2pi\_110397.fits} for CTI
corrections. For RDD corrections, we need to look for the frame data
obtained near to the epoch of our observation. Therefore, we picked up
{\tt rdd9812\_s[0 or 1]c[0, 1, 2, or 3]m2\_t0\_60.fits} for RDD
corrections retrieved from the Goddard Space Flight Center anonymous ftp
site.

\section{Spectrum Analysis}

The GIS observation shows a shell-like structure similar to that
obtained with ROSAT.  Its X-ray spectrum is almost uniform in spectral
shape over the remnant, consisting of two components: a power-law
component with a photon index, $\Gamma$, of $\sim 2.5$ and thin thermal
emission of electron temperature, $kT_{\rm e}$, of $0.07-0.21$ keV.  The
X-ray spectrum from the entire region is shown in figure 3.  RX
J0852$-$4622 overlaps with the Vela SNR that emits thin thermal
emission with $kT_{\rm e}$ of about 0.17 keV.  This value is very close
to that of the low temperature component seen in RX J0852$-$4622.
Furthermore, we find the intensity of this component to be very similar
to that appeared in the Vela SNR component around RX J0852$-$4622.  We
assume that the low temperature component seen in RX J0852$-$4622 is
background emission from the Vela SNR.  Therefore we can expect that the
emission from RX J0852$-$4622 is predominantly of the power-law type
originating from a shell-like region.

The SIS can cover only the central region of each GIS FOV.  Among them,
the spectrum of the bright north-west shell clearly shows features which
differ from those observed elsewhere as shown in figure 4.  We see a
clear emission line structure around 4 keV.  We see this structure only
in the eastern part of the north-west bright shell region.  Other
regions covered with the SIS show featureless spectra similar to that
obtained with the GIS for the entire remnant. For comparison, figure 5
shows the SIS spectrum extracted from the rectangular region shown with
the broken line in figure 1. There is no clear emission line structure
around 4 keV. This spectrum was well represented by a thin thermal
emission with $kT_{\rm e}$ of $0.10^{+0.01}_{-0.02}$ keV adding a
power-law spectrum with $\Gamma$ of $2.8^{+0.1}_{-0.2}$. Quoted errors
in this paper are at 90\% confidence level.

We fitted the SIS spectrum of the north-west shell with a two-component
model. First of all, we employed a low $kT_{\rm e}$ thin thermal
emission model with cosmic abundance ({\tt mekal}) in order to represent
the emission from the Vela SNR. We then added another component for the
emission from RX J0852$-$4622 described below. The neutral hydrogen
column density, $N_{\rm H}$, of two components were set to be equal as
one free parameter.

We applied a thermal bremsstrahlung model for the emission from RX
J0852$-$4622 proper, but we could not obtain an acceptable fit as shown
in model-1 of table 2. The big discrepancy between data and model is
found around 4 keV. We thus added a Gaussian component to reproduce the
structure around 4 keV (model-2). In this fitting, we fixed sigma of
Gaussian to be 0. Then, $\chi^2$ is improved by $\sim$ 10 compared with
the model-1 shown in table 2.  If we apply the F-test, the significance
level of an additional 4 keV line is $\sim$ 99\%. The line center energy
is $4.1 \pm 0.2$ keV and flux is $(5 \pm 2) \times 10^{-5}$ counts
s$^{-1}$ cm$^{-2}$ over the SIS FOV.

We then employed the thin thermal model with variable cosmic abundance
({\tt vmekal}) (model-3). Free parameters are $kT_{\rm e}$, the
normalization of two thermal emission components, the abundances of Si,
S, and Ca of the high $kT_{\rm e}$ component, and $N_{\rm
H}$. Abundances of the other elements are fixed to cosmic values. This
model is acceptable at 90\% confidence level. The Ca abundance is $8 \pm
5$ times larger than the cosmic value.  The absolute value is less
meaningful since it strongly depends on the applied model and other
model parameters. However, the relative abundances among the elements
are relatively robust. As shown in table 2, the Ca abundance is 11 times
and 40 times larger than those of Si and S, respectively.

If we employ a thin thermal model with nonequilibrium collisional
ionization (NEI) (Masai 1984; Masai 1994), $kT_{\rm e}$ increases while
the metal abundances do not change within the statistical uncertainties
as shown in table 2 (model-4).

\section{Discussion}

There are some candidates for the 4.1$\pm$0.2 keV emission line from an
astrophysical point of view.  If it is a characteristic X-ray line, it
must be either from neutral Sc-K or a helium-like Ca-K emission line.
From the cosmic abundance point of view, it is very implausible that it
comes from Sc.  If it comes from shock-heated matter, it is plausible
that it comes from Ca at high temperatures.  If this is not the case, it
can be either a red-shifted Fe-K line from an AGN or a cyclotron
emission line from a strongly magnetized neutron star.  Since the line
emission region is quite extended judging from the ASCA image, none of the
hypotheses of a point-like source is very likely.

The overabundance of Ca can be a result of an overabundance of $^{44}$Ti
which is produced only in the SN explosive nucleosynthesis process.  It
is mainly produced deep in the interior of the star, both in a type Ia
and type II SNe.  In the following we assume that the ejecta containing
$^{44}$Ti are uniformly expanding into ambient space.  In the early
phase of the SNR evolution, the major part of the emission does not come
from the shock heating process but comes from accelerated and
decelerating particles, which produce a power-law type X-ray spectrum.
In the north-west bright shell region, the ejecta might have recently
collided with an interstellar cloud forming shock-heated thermal
plasma, which is the source of the Ca emission line.

The expanding ejecta happen to hit the interstellar cloud at the
north-west shell.  The ejecta expanding along other directions should
also contain similar amounts of Ca but they may not yet been
shock-heated, so that just the power-law type spectrum prevails with no
emission lines.  We note that there is another bright shell region in
the south, which we did not observe in our ASCA observations.

Based on the emission line detected, we can estimate the total amount of
Ca.  The north west bright shell region is about 11$^\prime$ square
where we detected the emission line.  We assume that the depth of this
region is as equal to the width.  Assuming that the filling factor of
this region is unity, we find that the plasma density in this region is
$(0.9 \pm 0.1) (D/200{\rm\ pc})^{-0.5}$ H cm$^{-3}$ where $D$ is the
distance to the source, resulting a total mass to be $5.4 \times 10^{-3}
M_\odot$. Then, we can evaluate the total mass in all directions
assuming spherical symmetry.  In this way, we obtained the total amount
of Ca, $M_{\rm Ca}$, contained in RX J0852$-$4622 to be $1.1 \times
10^{-3} M_\odot (D/200 {\rm\ pc})^{2.5} (A/8)$, where $A$ is the
abundance of Ca.  We assume that all the $^{44}$Ca is in the shell
region but only that fraction expanding towards the north-west shell is
actually shock-heated and visible in the Ca line.

The major isotope of Ca on Earth is $^{40}$Ca while the fraction of
$^{44}$Ca is about $2 \times 10^{-2}$ (Anders \& Grevesse 1989). 
Assuming that the other isotopes of Ca are produced with the abundances
similar to the terrestrial values, $^{44}$Ca, the product of $^{44}$Ti,
should be heavily over-abundant, and we expect that almost all the Ca
detected must be $^{44}$Ca.

Based on the theoretical models of Nomoto et al. (1984) and Thielemann
et al. (1996), type II SN can produce Ca of 5$\times$10$^{-3}M_\odot$,
of which value slightly depends on the progenitor star mass, whereas
type I SN can produce $1.2 \times 10^{-2}\ M_\odot$.  Since we do not
know whether we have detected all Ca in RX J0852$-$4622, it is
inappropriate to compare our results with those of the model
calculations.  In both cases, the mass ratio of Ca to Si is less or
similar to that of the cosmic value.  These models cannot explain the
overabundance of Ca.  Whereas these models assume the point symmetric
explosion, asymmetric explosions can produce large metal abundance
anomalies (Nagataki et al. 1997), which may explain the overabundance of
Ca.  If we assume that most of Ca comes from $^{44}$Ti, we can estimate
the age, $t$, of RX J0852$-$4622 using the observed flux, $F$, of the
1157\,keV $\gamma$-rays.  We obtain t/$\tau\sb{1/2}$ = 16 $-$ log$_2$
(($F/3.8\times\ 10^{-5}$ photon cm$^{-2}$ s$^{-1}$) $(D/200 {\rm\
pc})^{1/2} (A / 8)$).  This value results in an upper limit of the age
of 970\,yrs and the lowest limit for the expansion velocity of 2900\,km
s$^{-1}$.  Since there occurs shock-heating associated with deceleration
of the expansion, the initial expansion velocity must have been higher
than this value.  If we assume that Ca in RX J0852$-$4622 has an isotope
population similar to that of the terrestrial value, the mass of
$^{44}$Ti is reduced by a factor of 50.  Then we obtain a lower limit of
the age of 630\,yrs and an upper limit of the expansion velocity of
4500\,km s$^{-1}$.

Model calculations of SNe show that the heavy elements are produced
relatively deep inside the progenitor star.  In type II SN, $^{44}$Ti is
produced very close to the mass-cut point, which is the critical
boundary separating the ejecta and the mass forming a central compact
remnant.  If the SN explosion occurs in a homologous fashion, $^{44}$Ti
and Ca are concentrated in a narrow layer.  We can expect that the
entire Ca is shock-heated simultaneously.  Based on the SN model
calculations the expanding velocity of the Ca dominated layer is a few
thousand km s$^{-1}$ (Shigeyama et al. 1988).  Therefore, the age of RX
J0852$-$4622 is likely to be closer to 970 yrs rather than 630 yrs.

\section{Conclusion}

We observed the SNR, RX J0852$-$4622, that is the $^{44}$Ti $\gamma$-ray
source.  It is towards the direction to the Vela SNR that is 250pc away
from us.  Based on the spectral analysis, the apparent spectrum from the
entire remnant consists of two components; one is a power-law component
and the other is thermal emission coming from the Vela SNR.

In the north-west bright shell, we have detected an emission line feature.
This feature is only seen by the SIS over a small
fraction of the remnant.  The line center is $4.1 \pm 0.2$ keV which is
interpreted as an emission line from highly ionized Ca.  We find that
the Ca abundance in this region is about $8 \pm 5$ times that of the cosmic 
value while the other elements are subcosmic.

If the Ca abundance anomaly mainly comes from the overabundance of
$^{44}$Ca that is produced from $^{44}$Ti, we can evaluate the age of
this source using the flux of the $^{44}$Ti $\gamma$-rays.  Since we
cannot distinguish among the Ca isotopes, the age estimate is between
630 yrs and 970 yrs.  If the Ca isotopes except $^{44}$Ca are produced
at cosmic abundance levels, the age is 970 yrs.

\vspace{1pc}\par
We thank {\sl ASCA\_ANL} and {\sl DISPLAY 45} developing team members,
especially Dr. Ishisaki.  We are grateful to all the members of ASCA
team for their contributions to the fabrication of the apparatus, the
operation of ASCA, and the data acquisition.  This research is partially
supported by the Sumitomo Foundation.

\clearpage

\section*{References}
\re Ahmad I, Bonino G., Castagnoli G.C., Fischer S.M., Kutschera W.,
 Paul M. 1998, Phys. Rev. Lett. 80, 2550
\re Anders E., Grevesse N. 1989, Geochim. Cosmochim. Acta, 53, 197
\re Aschenbach B. 1998, Nature, 396, 141
\re Aschenbach B., Iyudin A.F., Sch\"onfelder V., 1999, A\&A, 350, 997
\re Burgess C. P. Zuber K. 1999, astro-ph/9909010  
\re Cha R.N., Sembach K.R. Danks A.N. 1999, ApJ, 515, L25
\re Ishisaki Y. 1996, Ph.D. thesis of Univ. of Tokyo, ISAS RN 613
\re Iyudin A.F., Sch\"onfelder V., Bennett K., Bloemen H., Diehl R.,
Hermsen W., Lichti G.G., van der Meulen R.D. et al. 1998, Nature, 396, 142
\re Iyudin A.F., Diehl R., Bloemen H., Hermsen W.,
Lichti G.G., Morris D., Ryan J., Sch\"onfelder V. et al. 1994, A\&A, 284, L1
\re Koyama K., Petre R., Gotthelf E.V., Hwang U.
Matsuura M., Ozaki M., Holt S.S. 1995, Nature, 378, 255
\re Makishima K., Tashiro M., Ebisawa K., Ezawa H.,
Fukazawa Y., Gunji S., Hirayama M., Idesawa E. et al. 1996, PASJ 48, 171
\re Mewe R., Gronenschild E.H.B.M., van den Oord G.H.J.,
1985, A\&AS, 62, 197
\re Masai K., 1984, Ap\&SS 98, 267 
\re Masai K. 1994, Ap.J., 437, 770
\re Nagataki S., Hashimoto M., Sato K. Yamada S. 1997, ApJ, 486, 1026 
\re Nomoto K., Thielemann F.-K., Yokoi K., 1984, ApJ, 286, 644
\re Serlemitsos P.J., Jalota L., Soong Y., Kunieda
	H., Tawara Y., Tsusaka Y., Suzuki H., Sakima Y. et al. 1995, PASJ 47,
	105
\re Shigeyama T., Nomoto K., Hashimoto M. 1988, A\&Ap, 196, 141 
\re Tanaka Y., Inoue H., Holt S.S. 1994, PASJ 46, L37
\re The L.-S., Leising M.D., Clayton D.D., Johnson W.N., Kinzer R.L.,
Kurfess J.D., Strickman M.S., Jung G.V. et al. 1995, ApJ, 444, 244
\re Thielemann F.-K., Nomoto K., Hashimoto M., 1996, ApJ, 460, 408
\re Ueda Y., Takahashi T., Inoue H., Tsuru T., Sakano M.,
        Ishisaki Y., Ogasaka Y., Makishima K. et al. 1999, ApJ, 518, 656
\re Yamashita A., Dotani T., Bautz M., Crew G., Ezuka
H., Gendreau K., Kotani T., Mitsuda K. et al. 1997, IEEE Trans. Nuc.
Sci. 44, 847
\re Yamashita A., Dotani T., Ezuka H., Kawasaki M., Takahashi K.
1999, Nuclear Inst. Method A, 436, 68

\clearpage

\begin{fv}{1}{5cm}{The circular FOVs of the GIS are superimposed on the X-ray
intensity map obatained with ROSAT.  There are two bright shell regions
in the north and south.}
\end{fv}

\begin{fv}{2}{5cm}{X-ray surface brightness map obtained with the ASCA GIS.
The GIS image is smoothed with a Gaussian of
 $\sigma$=30$^{\prime\prime}$ after rebinning of 2$\times$2
 (30$^{\prime\prime}$ square) pixels.}
\end{fv}

\begin{fv}{3}{5cm}{The X-ray spectrum from the entire remnant obtained
with the GIS}
\end{fv}

\begin{fv}{4}{5cm}{The X-ray spectrum from the white square shown in
figure 1.  Solid lines represent the best fit curve of model-3 (see 
text).  Ca is overabundant by $8\pm 5$.}
\end{fv}

\begin{fv}{5}{5cm}{The X-ray spectrum from the rectangular region shown
 with the broken line in figure 1. Solid lines represent the best fit
 curve of the model of thin thermal emission adding a power-law
 spectrum.}
\end{fv}

\clearpage

\begin{table*}[t]
  \begin{center}
    Table~1. Observational Log\\
  \end{center}
  \vspace{6pt}
  \begin{tabular*}{\columnwidth}{@{\hspace{\tabcolsep}
  \extracolsep{\fill}}p{5pc}cccccc}
  \hline\hline\\[-6pt]
   Pointing ID & RA (J2000) & Decl.(J2000) &
   SIS chips & GIS bit assignment$^\ast$ & Exposure time$\dagger$ (ks) \\
   [4pt]\hline \\[-6pt]
   N1\dotfill & 8$^{\rm h}$49$^{\rm m}$ 18.$\!\!^{\rm s}$1 & $-$45$^{\rm d}$ 41$^\prime$ 0$^{\prime\prime}$  & S0C12, S1C30 & 8--6--6--5--0--6 & 9.3 \\
   N2\dotfill & 8$^{\rm h}$53$^{\rm m}$ 42.$\!\!^{\rm s}$4 & $-$45$^{\rm d}$ 32$^\prime$ 29$^{\prime\prime}$ & S0C01, S1C01 & 8--6--6--0--0--10 & 10.4 \\
   N3\dotfill & 8$^{\rm h}$55$^{\rm m}$ 31.$\!\!^{\rm s}$2 & $-$45$^{\rm d}$ 57$^\prime$ 12$^{\prime\prime}$ & S0C12, S1C30 & 8--6--6--5--0--6 & 13.7 \\
   N4\dotfill & 8$^{\rm h}$51$^{\rm m}$ 43.$\!\!^{\rm s}$2 & $-$46$^{\rm d}$ 7$^\prime$ 31$^{\prime\prime}$  & S0C12, S1C30 & 8--6--6--5--0--6 & 11.3 \\
   N5\dotfill & 8$^{\rm h}$47$^{\rm m}$ 47.$\!\!^{\rm s}$1 & $-$46$^{\rm d}$ 18$^\prime$ 5$^{\prime\prime}$  & S0C12, S1C30 & 8--6--6--5--0--6 & 9.8 \\
   N6\dotfill & 8$^{\rm h}$51$^{\rm m}$ 6.$\!\!^{\rm s}$6 & $-$46$^{\rm d}$ 46$^\prime$ 58$^{\prime\prime}$  & S0C12, S1C30 & 8--6--6--5--0--6 & 11.9 \\
   N7\dotfill & 8$^{\rm h}$54$^{\rm m}$ 50.$\!\!^{\rm s}$9 & $-$46$^{\rm d}$ 32$^\prime$ 17$^{\prime\prime}$ & S0C12, S1C30 & 8--6--6--5--0--6 & 9.8 \\
  \hline
  \end{tabular*}
  \vspace{6pt}\par\noindent
 $^\ast$ Each field stands for bit number of a pulse-height, x coordinate,
 y coordinate, rise-time, spread discrimination, and timing.\\
 $\dagger$ Exposure time of the GIS is shown.
\end{table*}

\clearpage

\begin{table*}[t]
  \begin{center}
   Table~2. Fitting results of the SIS spectrum at the north-west region.\\
  \end{center}
  \vspace{6pt}
  \begin{tabular*}{\columnwidth}{@{\hspace{\tabcolsep}
  \extracolsep{\fill}}p{4pc}cccccccc}
  \hline\hline\\[-6pt]
   model & low $kT_{\rm e}$ & high $kT_{\rm e}$
   & Si & S & Ca
   & $N_{\rm H}$ & $\chi^2$ (dof) \\
   & (keV) & (keV) & & & &  ($10^{22}$cm$^{-2}$) & \\
   [4pt]\hline \\[-6pt]
   model-1 & $(3.28^{+0.04}_{-0.66}) \times 10^{-2}$ & $2.0^{+0.4}_{-0.3}$
   & --- & --- & ---
   & $0.12^{+0.09}_{-0.05}$ & 99.3 (81) \\
   model-2 & $(3.29^{+0.04}_{-0.63}) \times 10^{-2}$ & 1.8$\pm$0.3
   & --- & --- & ---
   & $0.14^{+0.09}_{-0.06}$ & 89.2 (79) \\
   model-3 & $(4.25^{+0.70}_{-0.05}) \times 10^{-2}$ & 1.32$\pm$0.09
   & 0.7$\pm$0.4 & $\le$ 0.2 & 8$\pm$5
   & 1.1$\pm$0.1 & 87.2 (78)\\
   model-4$^\ast$ & $(5 \pm 1) \times 10^{-2}$ & $3.1^{+1.6}_{-0.8}$
   & $0.9^{+0.3}_{-0.2}$ & $\le$ 0.1 & $7^{+7}_{-5}$
   & $0.83\pm 0.1$ & 88.9 (77)\\
  \hline
  \end{tabular*}
  \vspace{6pt}\par\noindent
 Quoted errors are at 90\% confidence level. \\
 $^\ast$ Obtained ionization parameter is $10.8 \pm 0.1$.
\end{table*}

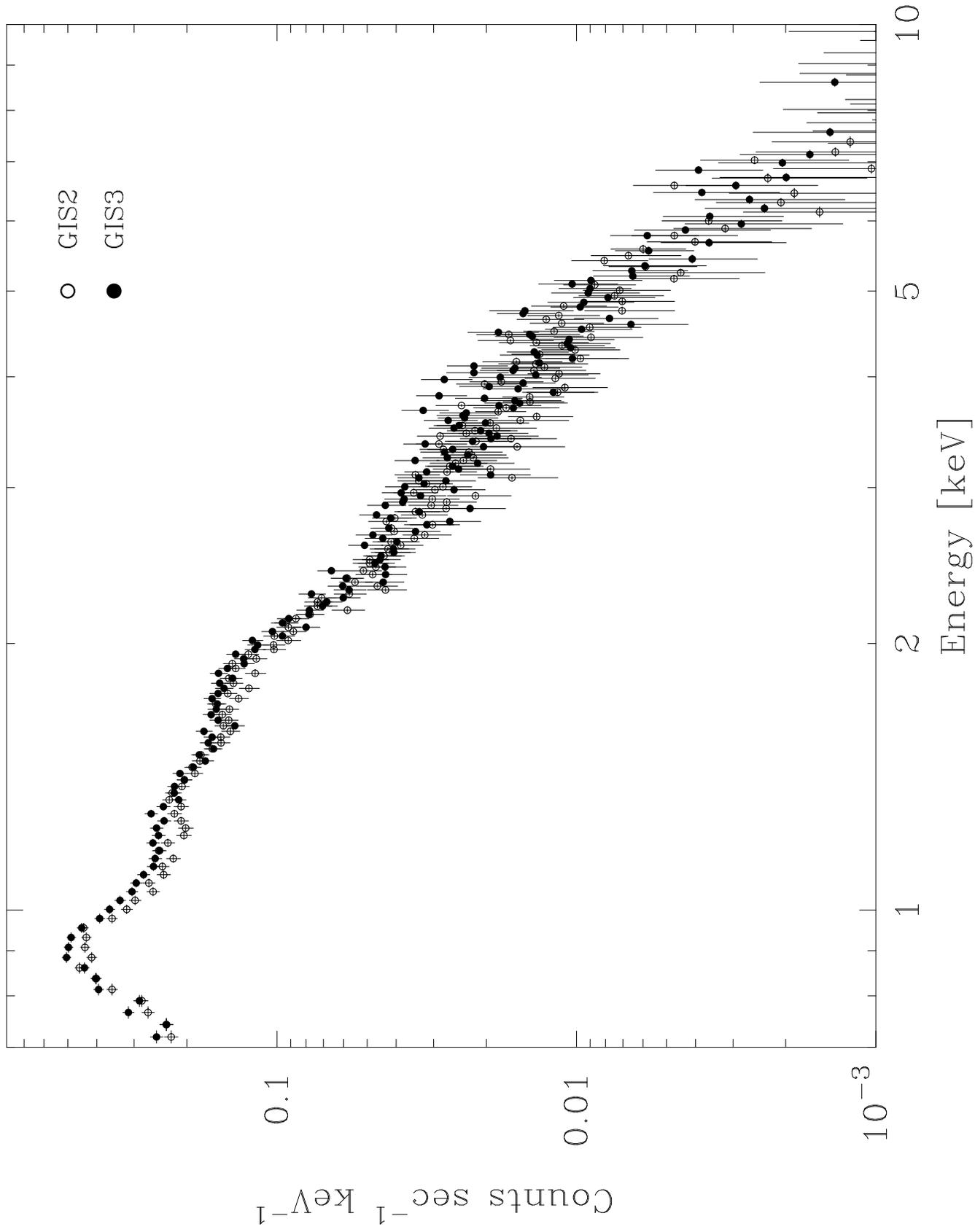
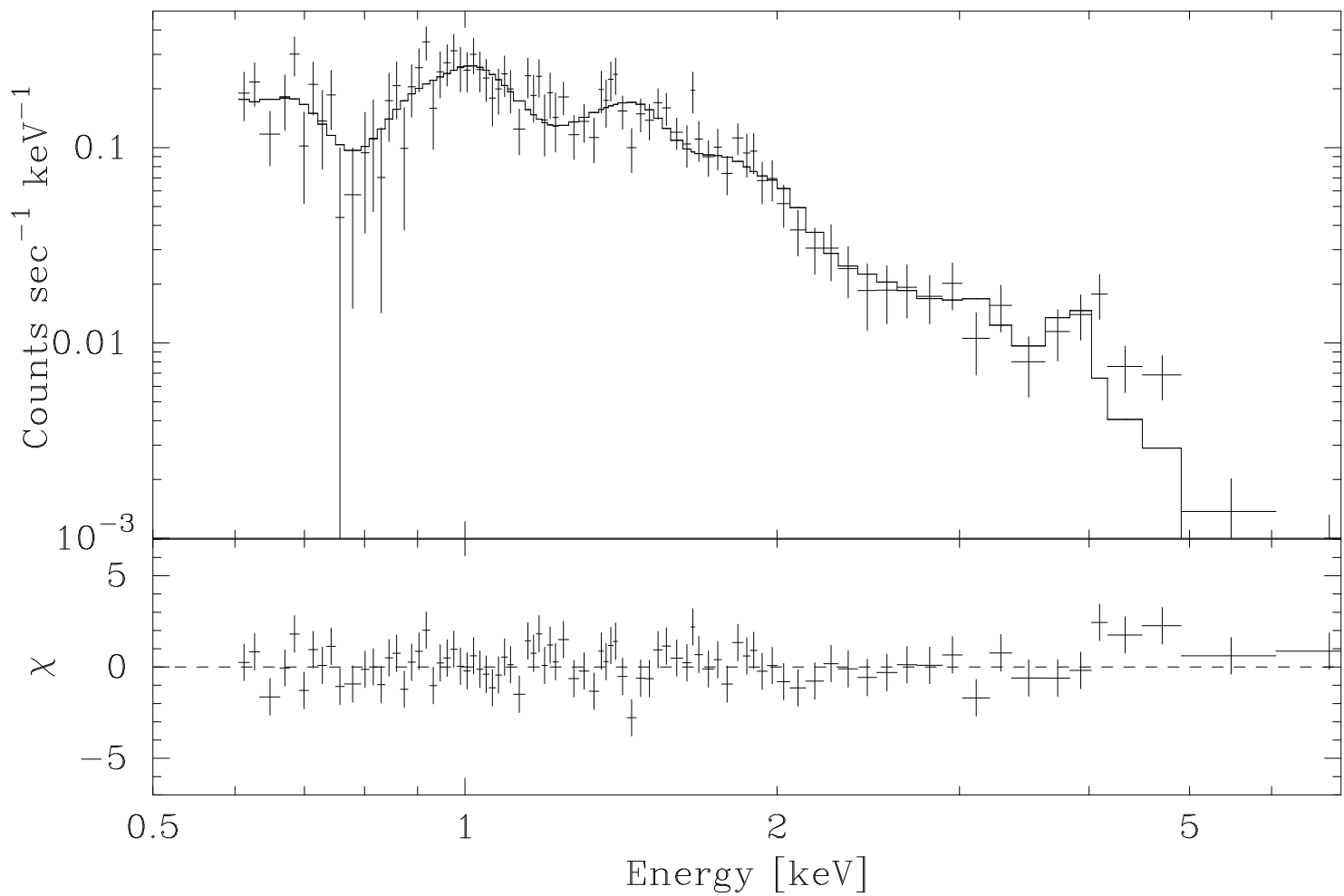
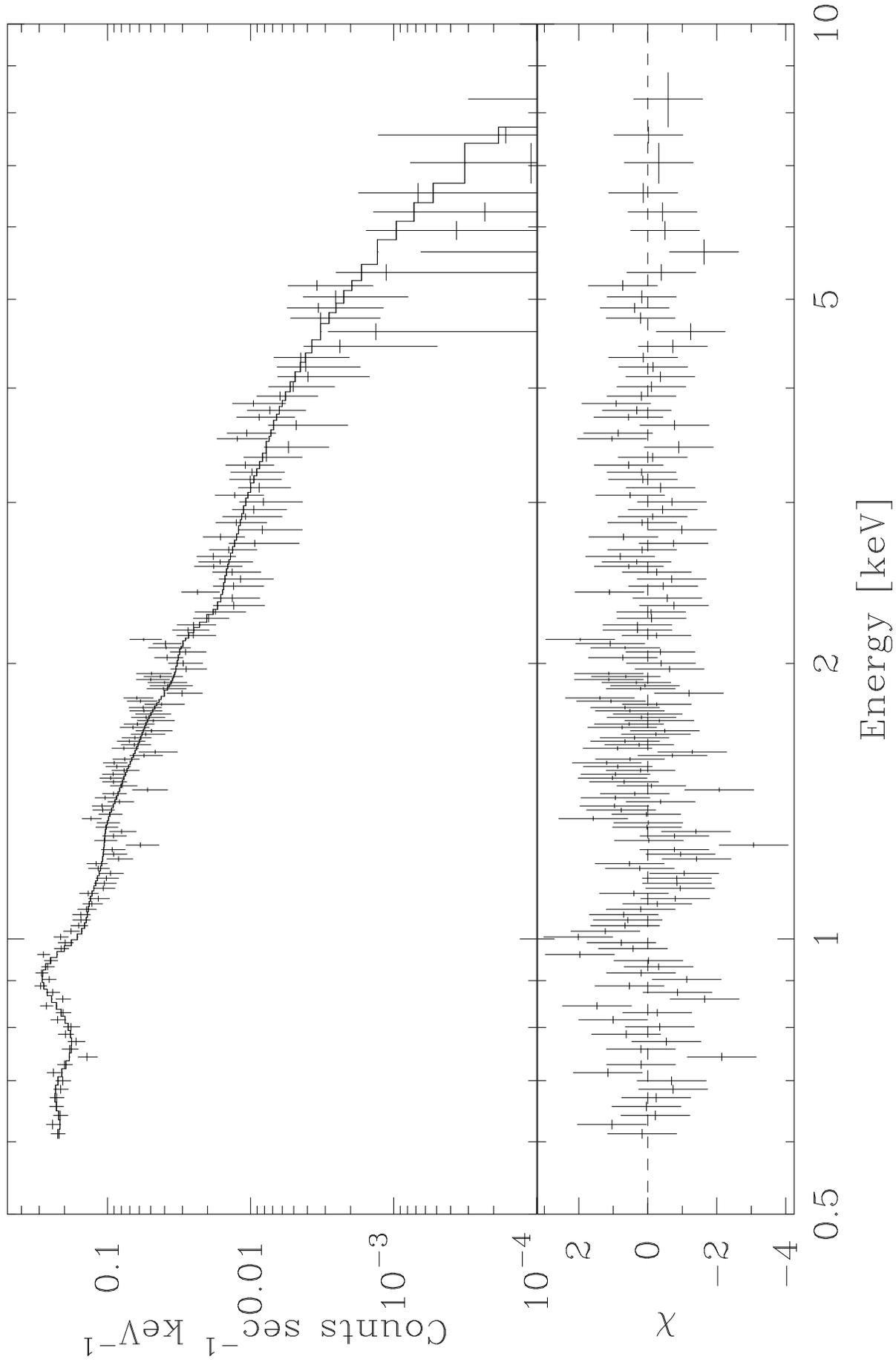
\end{document}